\documentclass[preview, 3pt]{elsarticle}
\usepackage{amssymb,amsmath,natbib}
\usepackage{graphicx,subfigure,epsfig}
\usepackage{dcolumn}

\usepackage{pgfplots}
\usepackage{tikz}
\usepackage{tikz-3dplot}
\usetikzlibrary{snakes}
\usetikzlibrary{calc,trees,positioning,arrows,chains,shapes.geometric,decorations.pathreplacing,decorations.pathmorphing,shapes,matrix,shapes.symbols}

\begin{document}

\title{\textbf{A non-local phase field model of Bohm's quantum potential}}

\author{\textbf{Roberto Mauri}\footnote{Author to whom correspondence should be addressed. Electronic mail:  roberto.mauri@unipi.it.}}

\address{Department of Civil and Industrial  Engineering, \\
Laboratory of Multiphase Reactive Flows, \\
Universit\`a di Pisa, I-56126 Pisa, Italy\\}

\begin{abstract}

Assuming that the free energy of a gas depends non-locally on the logarithm of its mass density, the body force in the resulting equation of motion consists of the sum of density gradient terms. Truncating this series after the second term, Bohm's quantum potential and the Madelung equation are identically obtained, showing explicitly that some of the hypotheses that led to the formulation of quantum mechanics admit a classical interpretation based on non-locality.

\end{abstract}

\maketitle

\section{Introduction} \label{sec I}

In 1927 Madelung \cite{Madelung 1927} showed that the Schrodinger equation for one-electron problems can be transformed into hydrodynamical equations. These equations describe the time evolution of the density and velocity fields of an ideal (i.e., non-viscous) gas subjected to the action of both classical and quantum potentials. This latter represents the novelty of the Madelung equation; it is generally referred to as the Bohm potential \cite{Bohm 1952a, Bohm 1952b}  and is a central concept of the de Broglie-Bohm formulation of quantum mechanics (see review in  \cite{BH 1989}). Accordingly, the Schrodinger equation can be interpreted as a Hamilton-Jacobi equation associated to the Madelung hydrodynamic equations, so that Madelung could conclude  \cite{Madelung 1927}   that there is thus a chance of erecting the quantum theory of the atom on this basis. In fact, the Madelung equations are the basis for a full description of quantum fluids  \cite{Barenghi 2016}.

Since then, many researchers have observed that, as the Bohm quantum potential derives from the Laplacian term of the Schrodinger equation, quantum effects must the consequence of some sort of non-local interaction. The non-local nature of quantum mechanics is further stressed by the fact that, due to the instantaneous nature of the collapse of the wavefunction, possible actions performed in a certain space region can induce immediate changes in distant regions   \cite{BH 1989, BG 2003}. Nevertheless, a clear explanation of how the Schrodinger equation derives from the non-locality of the force field is still missing.

In this work, we propose to fill this gap, starting from a non-local characterization of an ideal, i.e., non-viscous, fluid. Then, considering particle density as a phase field, we show that its evolution is described by the Madelung hydrodynamic equations.

\section{From non-locality to Schr\"{o}dinger equation} \label{sec II}

Consider an ideal fluid in isothermal conditions, that is a system composed of a large number of identical particles of mass m that do not interact with each other except for instantaneous thermalizing collisions. Its governing equations, expressing the conservation of mass and momentum, read \cite{deGrootMazur1962}:

\begin{eqnarray}  \label{1}
\frac{\partial \rho}{\partial t} + \nabla \cdot \left(  \rho \mathbf{v}  \right)  = 0,   \hspace{4 em} & \text{i.e.,}   & \hspace{3 em}   \frac{d \rho}{d t} = - \rho  \left(  \nabla \cdot \mathbf{v}  \right) ,
\\  \label{2}
\frac{\partial \left( \rho \mathbf{v} \right)  }{\partial t} + \nabla \cdot \left(  \rho \mathbf{v} \mathbf{v}   \right)  = \rho \mathbf{F},   \hspace{1 em}  & \text{ i.e.,} & \hspace{3 em}   \rho \frac{d \mathbf{v}}{d t} = \rho \mathbf{F},
 \end{eqnarray}
with  $ dA / dt = \partial A / \partial t + \mathbf{v} \cdot \nabla A   $   denoting the material derivative. Here, $ \rho $ is the density, $ \mathbf{v} $  is the fluid velocity, while $ \mathbf{F} $  is a force per unit mass, so that $ \rho \mathbf{F} $   is a body force, i.e., a force per unit volume. Note that $  \rho $  and $ \mathbf{v} $ are mean values, defined within an elementary point volume, assuming local equilibrium. In addition, $ \rho $ is proportional to the probability distribution to find a particle in a unit volume located in $ \mathbf{r} $ at time $t$. Now, considering that the probability distribution for a combination of two mutually independent subsystems is equal to the product of the distribution functions of the two subsystems separately, we reach the conclusion that $ \ln \rho $  must be a scalar and additive integral of motion, and therefore it must be proportional to the energy (see Landau et al. \cite{Landau 1980}. Therefore, we may assume that the free energy per particle mass, $ \widehat{f} \left( \mathbf{r}, t \right) $ , of a gas at constant temperature $ T $ has the following form:

\begin{equation} \label{3}
\widehat{f} \left( \mathbf{r}, t \right) = \frac{kT}{m} \int{ u \left( | \mathbf{r} - \mathbf{r}' |  \right) \ln{\rho} \left( \mathbf{r}', t \right) d^3 \mathbf{r}' } + \phi \left( \mathbf{r}, t \right),
\end{equation}
where $ \phi \left( \mathbf{r}, t \right) $  is the potential energy per unit mass resulting from the action of an external conservative force, while $ u \left( x \right) $  is an interaction kernel between particles located at a distance $ x = | \mathbf{r} - \mathbf{r}' | $, with the normalization condition,  $ \int{u \left( x \right) d^3 \mathbf{x} } = 1 $ .  Dropping for convenience the time dependence, this expression can be rewritten in the more convenient form,

\begin{equation} \label{3A}
\widehat{f} \left( \mathbf{r} \right) =  \widehat{f}_{th} \left( \mathbf{r} \right) + \frac{kT}{m} \int{ u \left( | \mathbf{r} - \mathbf{r}' |  \right) \left[ \ln{\rho} \left( \mathbf{r}' \right) - \ln{\rho} \left( \mathbf{r} \right) \right] d^3 \mathbf{r}' },
\end{equation}
where
\begin{equation} \label{4}
\widehat{f}_{th} \left( \mathbf{r} \right) = \frac{kT}{m} \ln \rho \left( \mathbf{r} \right)  + \phi \left( \mathbf{r} \right)
\end{equation}
is the thermodynamic part, corresponding to the free energy of a uniform ideal gas. Expanding in Taylor series,
\begin{equation} \label{5}
\ln \rho \left( \mathbf{r} + \mathbf{x} \right) = \ln \rho \left( \mathbf{r} \right) +  \mathbf{x} \cdot \nabla \ln \rho \left( \mathbf{r} \right) + \frac12 \nabla \nabla : \ln \rho \left( \mathbf{r} \right) + \cdots
\end{equation}
and truncating the series after the second term (there can be no $ \nabla \ln \rho $ term, due the isotropy of the fluid), we find,
\begin{equation} \label{6}
\widehat{f} \left( \mathbf{r} \right) =  \widehat{f}_{th} \left( \mathbf{r} \right) + \widehat{f}_{nl}^{(2)} \left( \mathbf{r} \right)
\end{equation}
where the leading-order correction term,
\begin{equation} \label{7}
\widehat{f}_{nl}^{(2)} \left( \mathbf{r} \right) = - \frac12 \frac{kT}{m} a^2 \nabla^2 \ln \rho \left( \mathbf{r} \right)
\end{equation}
is the non-local part, with
\begin{equation} \label{8}
 a^2 = - \int{ x^2 u \left( x \right) d^3 \mathbf{x} }
\end{equation}
denoting the square of a characteristic length, $a$. Note the negative sign in Eq. (\ref{8}), revealing that particles are assumed to attract each other.

Now we define a free energy per unit volume, $ \rho \widehat{f} $, so that the total free energy is given by the following functional:
\begin{equation} \label{9}
F \left[ \rho \left(  \mathbf{r} \right) \right] = \int_V { \rho \left( \widehat{f}_{th} - \frac12 \frac{kT}{m} a^2 \nabla^2 \ln \rho \right) d^3 \mathbf{r}  },
\end{equation}
where $ V $ is the total volume, that we assume to be infinite. Integrating by parts with the assumption that $ \rho \rightarrow 0 $  exponentially as $ r \rightarrow \infty $, find:
\begin{equation} \label{10}
F \left[ \rho \left(  \mathbf{r} \right) \right] = \int_V { f d^3 \mathbf{r} } = \int_V { \rho \left[ \widehat{f}_{th} + \frac12 \frac{kT}{m} a^2 \left( \nabla \ln \rho \right)^2  \right] d^3 \mathbf{r} },
\end{equation}
where $ f $ denotes the effective free energy per unit volume.

At equilibrium, the free energy functional (\ref{10}) is minimal, under mass conservation constraint, $ \int{ \rho d^3\mathbf{r} } = M \text{ constant} $. Therefore, applying the Euler-Lagrange equation, we see that the Legendre multiplier defines the following generalized chemical potential,
\begin{equation}\label{11}
    \mu = \frac{\delta f}{\delta \rho} = \frac{\partial f}{\partial \rho} - \nabla \cdot \frac{ \partial f}{\partial \nabla f} = \mu_{th} - \frac{kT}{m} a^2 \left[ \nabla^2 \ln \rho + \frac12 \left( \nabla \ln \rho \right)^2 \right].
\end{equation}
Here, the first term on the RHS denotes the thermodynamic chemical potential (i.e., Gibbs' free energy per unit mass), i.e.,
\begin{equation}\label{12}
    \mu_{th} = d \left( \rho \widehat{f}_{th} \right) / d \rho =  \widehat{f}_{th} - \widehat{v}  d \widehat{f}_{th} / d \widehat{v} = \widehat{f}_{th} + P \widehat{v},
\end{equation}
where $ \widehat{v} = \rho^{-1} $ and $ P = - d \widehat{f}_{th} / d \widehat{v} $ are the specific volume and the pressure, respectively.

The second term on the RHS of Eq. (\ref{11}) is the non-local contribution to the generalized chemical potential, i.e.,
\begin{equation}\label{13}
    \mu_{nl}^{(2)} = - \frac{kT}{m} a^2 \left[ \nabla^2 \ln \rho + \frac12 \left( \nabla \ln \rho \right)^2 \right]
\end{equation}
The equations of motion can be determined applying a variational principle \cite{LMS 2017} to derive the Euler Eq. (\ref{2}) for a compressible, inhomogeneous fluid where $ \mathbf{F} $ is a body force, driven by density gradients in the fluid. The same result can be obtained more heuristically by applying Noether's theorem \cite{AmFW 1998} or by the least action principle \cite{LT 1998}. At the end, we find Eqs. (\ref{1}) and (\ref{2}), with the following force per unit mass:
\begin{equation}\label{14}
    \mathbf{F} = - \nabla \mu = - \nabla \mu_{th} - \nabla \mu_{nl}^{(2)}.
\end{equation}
Let us consider the two terms on the RHS separately. On one hand we have:
\begin{equation}\label{15}
\mathbf{F}_{th} = - \nabla \mu_{th}, \hspace{2 em} \text{with} \hspace{2 em} \mu_{th} = \frac{kT}{m} \left( \ln \rho + 1 \right) + \phi,
\end{equation}
and therefore, considering that $ P = \frac{kT}{m} \rho $,
\begin{equation}\label{16}
\rho \mathbf{F}_{th} = - \rho \nabla \mu_{th} = - \nabla P - \rho \nabla \phi.
\end{equation}
On the other hand, from Eqs. (\ref{13}) and (\ref{14}), we define a non-local reversible body force, which is usually referred to as the Korteweg force:
\begin{equation}\label{17}
\mathbf{F}_{nl}^{(2)} = - \nabla \mu_{nl}^{(2)} = \frac{kT}{m} a^2 \nabla \left[ \nabla^2 \ln \rho + \frac12 \left( \nabla \ln \rho \right)^2  \right].
\end{equation}
The non-local potential (\ref{13}) can be expressed in the following equivalent form:
\begin{equation}\label{18}
 \mu_{nl}^{(2)} = - 2 \dfrac{kT}{m} a^2 \frac{\nabla^2 \sqrt{\rho}}{\sqrt{\rho}}.
\end{equation}
In fact:
\begin{eqnarray} \nonumber
 \nabla^2 \ln \rho + \frac12 \left( \nabla \ln \rho \right)^2 = \frac{1}{\sqrt{\rho}} \left[ \sqrt{\rho} \nabla^2 \ln \rho + \frac{\nabla \rho  \cdot \nabla \ln \rho}{2 \sqrt{\rho} } \right] = \\  \label{19}
 = \frac{1}{\sqrt{\rho}} \nabla \cdot \left( \sqrt{\rho} \nabla \ln \rho \right) = \frac{1}{\sqrt{\rho}} \nabla \cdot \left( \frac{\nabla \rho}{\sqrt{\rho}} \right) = 2 \frac{ \nabla^2 \sqrt{\rho} }{\sqrt{\rho}}.
\end{eqnarray}
When $a$ is the thermal de Broglie wavelength \cite{GS 2020},
\begin{equation}\label{20}
    a = \frac{\hbar}{\sqrt{4 m kT}},
\end{equation}
then $ \mu_{nl} $ reduces to Bohm's quantum potential \cite{Bohm 1952a, Bohm 1952b}, i.e.,
\begin{equation}\label{21}
 \mu_{nl}^{(2)} = Q = - \frac{\hbar^2}{2 m^2} \frac{\nabla^2 \sqrt{\rho}}{\sqrt{\rho}}.
\end{equation}
Finally, summarizing, from Eq. (\ref{2}), the force balance per unit mass yields the Madelung equation:
\begin{equation}\label{22}
    \frac{\partial \mathbf{v}}{\partial t} + \mathbf{v} \cdot \nabla \mathbf{v} = \mathbf{F} = - \nabla V - \nabla Q,
\end{equation}
with
\begin{equation}\label{22A}
    V = \widehat{f}_{th} = \frac{kT}{m} \ln \rho + \phi, \hspace{1 em} \text{and} \hspace{1 em} \nabla V = \nabla P / \rho + \nabla \phi.
\end{equation}
Here, $ V $ denotes the thermodynamic potential energy (\ref{4}), while $ Q = \mu_{nl} $ is its non-local counterpart, that is the Bohm potential.

Since the fluid is frictionless, the flow is irrotational (i.e., with zero curl), and the fluid velocity can be expressed as the gradient of a real scalar function, namely the velocity potential, $ S $, that is,
\begin{equation}\label{23}
    \mathbf{v} = \nabla S.
\end{equation}
Substituting (\ref{23}) into (\ref{22}) we see that $ S $ satisfies the following Hamilton-Jacobi equation, describing the motion of a particle of mass $ m $ in a conservative force field, characterized by a classical potential energy, $ V $, and the quantum potential field, $ Q $ (see Eqs. (\ref{21}) and (\ref{22})),
\begin{equation}\label{24A}
    \frac{\partial S}{\partial t} + \frac12 \left( \nabla S \right)^2 + V - \frac{\hbar^2}{2 m^2} \frac{\nabla^2 \sqrt{\rho}}{\sqrt{\rho}}.
\end{equation}
This equation, together with the continuity equation (\ref{1}),
\begin{equation}  \label{24B}
\frac{\partial \rho}{\partial t} + \nabla \cdot \left(  \rho \nabla S  \right)  = 0,
 \end{equation}
can be obtained using the Madelung transform,
\begin{equation}\label{25}
    \psi = \sqrt{\rho} e^{iS},
\end{equation}
and imposing that the complex function $ \psi $  satisfies the Schr\"{o}dinger equation,
\begin{equation}\label{26}
    i \hbar \frac{\partial \psi}{\partial t} = - \frac{\hbar^2}{2 m} \nabla^2 \psi + V \psi,
\end{equation}
with  $ \frac{\hbar^2}{2 m} = 2 kT a^2 $.  Eq. (\ref{22}) can also be written as a Bernoulli equation as follows:
\begin{equation}\label{27}
    m \frac{\partial \mathbf{v}}{\partial t} = - \nabla E = - \nabla \left(  \dfrac12 m v^2 + mV + mQ \right),
\end{equation}
where the RHS denotes the gradient of the total energy, $ E $, that is the sum of kinetic energy, potential energy, and quantum potential energy. With some manipulation, we can write Eq. (\ref{27}) in the following equivalent form,
\begin{equation}\label{28}
    \rho \left( \frac{\partial \mathbf{v}}{\partial t} + \mathbf{v} \cdot \nabla \mathbf{v} \right) = \rho \mathbf{F} = - \rho \nabla V - \nabla \cdot \mathbf{P}_{Q}, \hspace{1 em} \mathbf{P}_{Q} = - \frac{\hbar^2}{4 m} \rho \nabla \nabla \ln \rho
\end{equation}
with $ \mathbf{P}_{Q} $ denoting the quantum stress tensor. Finally, it must be noted that the non-local term is not a small correction to its thermodynamic counterpart, as in Eq. (\ref{17}) $ \nabla^2 \rho \approx \rho / a^2 $,  so that the non-local potential $ Q $ is not necessarily smaller than the local, $ \mathcal{O}(kT/m) $  term.

Equation (\ref{28}) (or, equivalently, Eqs. (\ref{22}) and (\ref{28})), together with Eq. (\ref{1}), are known as the superfluid hydrodynamic equations \cite{Barenghi 2016}.

\section{Higher-order terms}\label{sec III}

Considering the higher-order terms in the expansion (\ref{5}), we obtain the following expansion for the free energy,
\begin{equation}\label{29}
    \widehat{f} = \frac{kT}{m} \sum_{n=0}^{\infty} (-1)^n \frac{c_{2n}}{(2n)!} a^{2n} \nabla^{2n} \ln \rho,
\end{equation}
where $ c_0 = 1 $ and  $ c_2 = 1 $  (this is the definition (\ref{8}) of $ a^2 $, while,
\begin{equation}\label{30}
    c_{2n} = (-1)^n \frac{1}{a^{2n}} \int_V { x^{2n} u \left( x \right) d^3 \mathbf{x} }; \hspace{0.5 em} n = 2, 3, \cdots
\end{equation}
are the normalized central moments of the $ u (x) $  distribution.

The total free energy then becomes, after integrating by parts:
\begin{eqnarray}\label{31}
    F \left[ \rho \left( \mathbf{r} \right) \right] = \int_V f d^3 \mathbf{r} = \frac{kT}{m} \sum_{n=0}^{\infty} (-1)^n \frac{c_{2n}}{(2n)!} a^{2n} \int_V { \rho \left( \nabla^{2n} \ln \rho \right) d^3 \mathbf{r}  } = \\  = F_{th} \left( \mathbf{r} \right) - \frac{kT}{m} \sum_{n=1}^{\infty} (-1)^n \frac{c_{2n}}{(2n)!} a^{2n} \int_V { \nabla \ln \rho \cdot  \left( \nabla \nabla^{2n-2} \rho \right) d^3 \mathbf{r}  },
\end{eqnarray}
where $ F_{th} = \frac{kT}{m} \int { \rho \ln \rho  d^3 \mathbf{r} } $.  Naturally, truncating the series (\ref{31}) after $ n=1 $, we find Eqs. (\ref{9}) and (\ref{10}). Now, minimizing this functional with the constraint of mass conservation we obtain:
\begin{equation}\label{32}
    \mu = \frac{\delta f}{\delta \rho} = \sum_{m=0}^{\infty} \frac{\left( -1 \right)^m}{m!} \nabla^m \frac{\partial f}{\partial \nabla^m \rho} = \mu_{th} + \sum_{n=1}^{\infty} \mu_{nl}^{(2n)},
\end{equation}
where $ \mu_{th} $ is the thermodynamic chemical potential (\ref{12}), while the non-local potentials $ \mu_{nl}^{(2n)} $  read:
\begin{equation}\label{33}
    \mu_{nl}^{(2n)} = \frac{kT}{m} (-1)^n \frac{c_{2n}}{(2n)!} a^{2n} \left[ \frac{1}{\rho} \nabla^{2n} \rho + \nabla^{2n} \ln \rho \right].
\end{equation}
In particular, $ c_2 = 1 $  and $ \mu_{nl}^{(2)} = Q $. Finally, the body force appearing in the Bernoulli equation (\ref{2}) is again equal to the expression (\ref{14}), with the Korteweg force now being equal to the sum of the gradients of the non-local potential (\ref{33}),
\begin{equation}\label{34}
    \frac{\partial \mathbf{v}}{\partial t} + \mathbf{v} \cdot \nabla \mathbf{v} = \mathbf{F} = - \nabla V - \nabla Q - \sum_{n=2}^{\infty} \nabla \mu_{nl}^{(2n)}.
\end{equation}
The last term on the RHS of Eq. (\ref{34}) represents a presumably small correction to the Madelung hydrodynamic equation (\ref{22}).

\section{Comments and conclusions}

In this work, Bohm's quantum potential is derived assuming non-local effects, namely a non-local dependence of the free energy of the system on the logarithm of its mass density through an interaction kernel, $  u(x) $. This implies that the body force in the Madelung hydrodynamic equations of quantum fluids consists of a density gradient expansion (see Eq. (\ref{34})), truncated after the second term. Clearly, that means that considering all the hierarchy of density gradient terms, the interaction kernel $ u(x) $ leads to very small corrections to the Schr\"{o}dinger equation. For example, assume that $ u(x) $ consists of a hard-core, short-range repulsion term of range $ d $ and a long-range attractive, exponentially decaying potential of range $ \ell $, i.e.,
\begin{equation}\label{35}
u \left( x \right) = \left\{
    \begin{split}
    n_0 \hspace{3 em} &  (x < d) \\
      - n_1 \, exp(- \frac{x}{\ell}) \hspace{1 em}   &  (x>d)
    \end{split}
    \right.
\end{equation}
in the limit $ d/\ell = \mathcal{O} (\epsilon) $ , with $ \epsilon \ll 1 $, assuming that $ n_0 d^3 \approx  n_1 \ell^3 = \mathcal{O} (\epsilon^{-1}) $, the normalization condition, $ \int { u(x) d^3 \mathbf{x} } = 1 $ , leads to $ n_0 = 9 n_1 \ell^3 / d^3 $, while Eq. (\ref{8}) yields: $ (a/\ell)^3 = 96 \pi (n_1 \ell^3) $, showing that $ a/\ell = n_0 d^3 = \mathcal{O} (\epsilon^{-1/2}) $. In general, Eq. (\ref{30}) gives:
\begin{equation}\label{36}
    c_{2n} = (-1)^{n+1} \frac{\left( 2n+2 \right)!}{24} \, \left( \frac{\ell}{a} \right)^{2(n-1)},  \hspace{1 em} n=1,2,\cdots,
\end{equation}
showing that, as $ \ell \ll a $,  the moments decrease as $ n $ increases. In particular, we confirm that $ c_2 = 1 $, while $ c_4 = - 30 (\ell/a)^2 = \mathcal{O} (\epsilon) $, $ c_6 = 1680 (\ell/a)^4 = \mathcal{O} (\epsilon^2)  $ , etc. Consequently, assuming that the interaction kernel is given by Eq. (\ref{35}), we see that the additional terms, correcting the Madelung equation (\ref{24A}), are very small.

Concluding, we have shown that some of the hypotheses that led to the formulation of quantum mechanics admit a simple classical interpretation, based on non-locality, so that quantum mechanics could be interpreted as a sort of mean field theory, that is a coarse-grained approximation of a more microscopic physical reality. What remains to be seen is the physical nature of the underlying non-local interaction field (see discussion by L. Smolin \cite{Smolin 2016}). In fact, although this hypothesis is not used in the mathematical development of the model, it is essential for its physical understanding, analogously to the background field hypothesis leading to the formulation of stochastic mechanics (see Nelson \cite{Nelson 1985}).

\end{document}